\documentclass[conference]{IEEEtran}
\IEEEoverridecommandlockouts

\usepackage{cite}
\usepackage{amsmath,amssymb,amsfonts}
\usepackage{algorithmic}
\usepackage{graphicx}
\usepackage{textcomp}
\usepackage{xcolor}
\usepackage{booktabs}
\usepackage{multirow}
\usepackage{multicol}
\usepackage{makecell}

\usepackage{amsthm}
\usepackage{amsmath}
\usepackage{amssymb}

\usepackage{comment}

\usepackage{amsfonts}

\usepackage{fancyvrb}
\usepackage{tabularx}

\usepackage[most]{tcolorbox}

\newtcolorbox{promptbox}{
  colback=gray!5,
  colframe=gray!50,
  boxrule=0.4pt,
  arc=2pt,
  left=6pt, right=6pt, top=4pt, bottom=4pt,
  fontupper=\small\ttfamily,
  breakable
}

\def\BibTeX{{\rm B\kern-.05em{\sc i\kern-.025em b}\kern-.08em
    T\kern-.1667em\lower.7ex\hbox{E}\kern-.125emX}}

\usepackage{wasysym}

\usepackage{dsfont}

\begin{document}

\title{Large Audio Language Models for Spoofing-Aware Speaker Verification}



\author{
\centering
\begin{tabular}{ccc}

\begin{minipage}[t]{0.31\textwidth}
\centering
\textbf{Sofya Savelyeva}\\
\textit{Applied AI Institute}
\end{minipage}
&
\begin{minipage}[t]{0.31\textwidth}
\centering
\textbf{Mariia Perunova}\\
\textit{MIRAI}
\end{minipage}
&
\begin{minipage}[t]{0.31\textwidth}
\centering
\textbf{Evgeny Kushnir}\\
\textit{AXXX, HSE}\\
\textit{Applied AI Institute}
\end{minipage}

\\[2.9em]

\begin{minipage}[t]{0.31\textwidth}
\centering
\textbf{Artem Dvirniak}\\
\textit{MIRAI}
\end{minipage}
&
\begin{minipage}[t]{0.31\textwidth}
\centering
\textbf{Dmitrii Korzh}\\
\textit{AXXX, MTUCI}\\
\texttt{d.s.korzh@mtuci.ru}
\end{minipage}
&
\begin{minipage}[t]{0.31\textwidth}
\centering
\textbf{Oleg Y. Rogov}\\
\textit{AXXX, Applied AI Institute}\\
\textit{MTUCI}
\end{minipage}

\end{tabular}
}

\maketitle

\begin{abstract}
Recent advances in text-to-speech and voice cloning make high-quality spoofing inexpensive and scalable, threatening voice authentication systems, especially automatic speaker verification (ASV). Existing defenses mainly address this threat through binary countermeasures (CMs) for deepfake detection or spoofing-aware speaker verification (SASV), where current systems are dominated by modular ASV-CM fusion and cascaded pipelines. Although large audio language models (LALMs) have shown promise on related audio tasks, including CM and ASV, their use for SASV remains unexplored, despite their capacity to produce natural-language rationales for auditing and robustness beyond discriminative predictions. This work systematically evaluates LALMs for SASV against conventional pipelines under zero-shot prompting, supervised adaptation, reasoning-oriented training, and reinforcement-learning-based optimization. Our results show that pretrained LALMs are near chance in the zero-shot setting, confirming that they are not natively suited to SASV, but that task-specific adaptation closes this gap. We further find that competitive SASV performance can be achieved through several distinct routes. These findings position LALMs as a promising and auditable foundation for unified SASV, while clarifying where conventional cascade systems still lead.
\end{abstract}


\section{Introduction}

Recent advances in text-to-speech (TTS) and voice cloning have substantially reduced the cost of producing realistic synthetic speech~\cite{casanova2024xtts,du2025cosyvoice}. This progress creates a direct threat to voice-based authentication, remote identity verification, and speaker-dependent human-machine interaction, where an attacker may attempt to synthesize speech that is both natural-sounding and attributable to a claimed speaker.

Most existing defenses treat this threat as binary speech deepfake detection (SDD), or countermeasure (CM) modeling, classifying audio as spoof (deepfake) or bona fide (genuine)~\cite{yi2023add,wang2024asvspoof}. Yet SDD robustness remains limited under distribution shifts from unseen spoofing algorithms, recording conditions, speakers, and post-processing. Spoofing-aware speaker verification (SASV) instead requires a joint decision-accept bona fide target trials, reject bona fide impostor trials, and reject spoofed trials~\cite{jung2022sasv}. By leveraging comparisons with enrollment audio, SASV partially mitigates the poor generalization of binary CM classifiers to new generators.

Although SASV is now an active research area, progress has been driven mainly by increasingly sophisticated integration of common automatic speaker verification (ASV)~\cite{desplanques2020ecapa} and CM models.
Prior work has shown that score-level fusion of CM and ASV scores can substantially reduce the vulnerability of ASV systems to spoofed trials~\cite{shim2022baseline}, outperforming the neural fusion of embeddings obtained through ASV and CM. Subsequent methods extend this paradigm through further embedding fusion, cascaded architectures, nonlinear calibration backends, task-aligned a-DCF optimization, and integrated discriminative formulations~\cite{heo2022two,wang2022dku,kurnaz2026joint,tan2026integrated}. Nonetheless, the primary outcome comes from strong CM and ASV models.

Large audio language models (LALMs)~\cite{tang2023salmonn} are gaining popularity for adaptation to specific applications. They can process audio within an instruction-following framework and produce structured natural-language evidence, offering reasoning capabilities to improve quality and interpretability. Namely, Chain-of-thought (CoT) prompting and training~\cite{guo2025deepseek} can elicit intermediate rationales and have been explored in such tasks as audio question answering~\cite{yang2025sakura} and audio quality assessment~\cite{wang2025speechllm}. Recent studies have also applied LALMs to SDD~\cite{wang2025speechllm,xu2026holiantispoof,chen2026audio}, while separate work investigates their use for speaker verification~\cite{ren2025can,nam2026speakerllm}. The evidence is mixed: LALMs generally perform poorly in zero-shot and few-shot speaker verification and SDD, yet supervised adaptation and reasoning-oriented training can improve their performance, though it is not guaranteed to outperform conventional models.

This raises a natural question: can LALMs be useful for SASV, can they provide acceptable verification performance, and can they provide adequate reasoning over their predictions? We address this question through a systematic study of LALMs for SASV. We formulate SASV not only as a discriminative task (accept/reject or target/non-target/spoof), but also as a pairwise audio reasoning task over enrollment and trial utterances, and compare prompting, supervised fine-tuning, reasoning supervision, and GRPO-based optimization. Our study asks whether LALMs improve over conventional pipelines, what helps to achieve better performance, whether reasoning supervision improves performance, and whether GRPO can provide additional gains. 

\noindent\textbf{Our contributions might be summarized as follows:}
\begin{itemize}
    \item We provide the first systematic evaluation of LALMs for spoofing-aware speaker verification.
    \item We propose several methods to improve LALMs' hard-label verification capabilities by combining task-specific losses to achieve a better CM-ASV trade-off. We demonstrate that LALMs can be a plausible solution,  outperforming strong baselines.
    \item We propose and evaluate chain-of-thought fine-tuning strategies and GRPO-based optimization for the SASV, using cold-start traces from large-scale LALM, conventional features, and human annotations.
\end{itemize}

\section{Related Work}

Recent evaluation campaigns have further broadened the scope of audio deepfake detection beyond the conventional ADD~\cite{yi2023add}  and ASVspoof~\cite{wang2024asvspoof} protocols. In particular, WildSpoof~\cite{wu2026wildspoof} emphasizes in-the-wild TTS generation and SASV using predefined audio sets and generated to bypass the guards samples, AT-ADD~\cite{xie2026add} evaluates robust speech and all-type audio deepfake detection under unseen generation methods and realistic distortions, RADAR~\cite{luong2026radar} targets media-transformation robustness, and ESDD2~\cite{zhang2026overview} extends the task to environmental and component-level audio manipulations. Existing SDD systems primarily improve empirical performance through architectural and training refinements, such as graph representations over spatial and temporal audio features~\cite{jung2022aasist}, augmentations and optimizers~\cite{tak2022rawboost,ding2023samo}, and primarily strong audio encoders, including task-specific front-ends such as learnable sinc-convolutions~\cite{tak2021end} and self-supervised learning (SSL)~\cite{tak2022automatic,zhu2026alethia} audio encoders.

SASV-2022~\cite{jung2022sasv,shim2022baseline} was one of the first major challenges devoted to the SASV. While most baselines and proposed methods use various score fusion strategies from the SDD countermeasure and general ASV models, end-to-end solutions remain limited. In~\cite{Alenin2022A}, the authors first train the ASV model and then additionally train the CM subnetwork, which processes intermediate features of the frozen ASV model. In~\cite{zeng2024spoofing}, authors study spoofing, channel, and domain mismatch with a specialized discriminative dual-path model. Their results stress the need for beyond-clean SASV evaluation, but the method remains a specialized discriminative system.

Following recent WildSpoof~\cite{wu2026wildspoof}, leading systems have improved robustness using stronger SSL frontends, domain-shift augmentation, codec-residual preprocessing, and calibrated ASV-CM fusion, yet their relative performance varies across SpoofCeleb~\cite{jung2025spoofceleb}, ASVspoof~5, SASV~2022, and newly generated TTS-track attacks. The BUT submission~\cite{peng2025but} combines speech-specific SSL encoders such as WavLM~\cite{chen2022wavlm} with general-audio encoders such as Dasheng~\cite{dinkel2024scaling}, uses Multi-Head Factorized Attention to aggregate SSL layers, and applies distribution uncertainty augmentation to improve robustness to unseen vocoders and recording conditions. Thus, current SASV systems remain largely discriminative and score-centric specialized pipelines. They improve robustness empirically, but they do not expose grounded evidence for the joint speaker-authenticity decision. This is the gap addressed by our work.



In parallel, recent studies investigate LALMs for speaker verification and SDD. In~\cite{ren2025can}, authors reformulate text-independent and text-dependent speaker verification as audio question answering over enrollment-trial pairs. They show that LALMs have limited zero-shot ASV ability and are sensitive to prompting, but Low-Rank Adaptation (LoRA)~\cite{hu2022lora} fine-tuning with hard-pair sampling substantially improves performance and can approach conventional ECAPA-TDNN~\cite{desplanques2020ecapa} or cascaded systems. However, these studies do not address joint speaker-authenticity decisions required by SASV.
SpeechLLM~\cite{wang2025speechllm} focuses on interpretable audio quality assessment and also reports hard-label SDD performance. HoliAntiSpoof~\cite{xu2026holiantispoof} uses LALMs for joint reasoning over spoofing mechanisms, temporal localization, and semantic artifacts. It improves hard-label performance and interpretability over binary classifiers, but does not evaluate SDD reasoning or SASV performance, requires task-specific retraining, and reports no ASVspoof5 results. Closest to our work, \cite{chen2026audio} builds a CoT reasoning dataset for ADD and tests its impact on detection. The model feeds WavLM-Plus embeddings, QFormer-projected~\cite{jiang2024qformer}, into Llama 3.2-1B-Instruct~\cite{grattafiori2024llama} with prompts listing acoustic cues such as $F_0$, pitch, and formants. It reasons over a bona fide reference and a trial utterance. However, it is evaluated mainly on ASVspoof~19 and private data, and reports weak ASV/SASV performance of about $75\%$ accuracy, which the authors attribute to the need for stronger speaker-centric supervision and dataset design.

\section{Methodology}


We formulate SASV as a three-way decision problem over an enrollment utterance and a trial utterance. Given an enrollment utterance with a claimed identity and a trial (query) utterance presented for verification, the system predicts one of three labels: target (verified genuine/bona fide speech from the enrolled speaker), non-target (human impostor speech), or spoof (synthetic or converted speech from any speaker). In contrast to cascaded ASV-CM pipelines that optimize speaker verification and spoofing countermeasure objectives separately and then fuse their scores post hoc, we train a single end-to-end model for the joint SASV objective. In this work, \textbf{we address the following research questions:}
\begin{description}
  \item[\textbf{RQ1}] Can pretrained LALMs be adapted to perform spoofing-aware speaker
  verification?

  \item[\textbf{RQ2}] How do LALM-based SASV systems perform compared to strong conventional competitors?

  \item[\textbf{RQ3}] Which adaptation and training strategies can help to improve performance and balancing the trade-off between speaker discrimination and spoof detection?

  \item[\textbf{RQ4}] Can explicit reasoning supervision improve the robustness and interpretability of LALM-based SASV beyond direct decision-level fine-tuning?
\end{description}
The remainder of this section describes the methodology behind the experiments designed
to answer these questions.

\subsection{Model Adaptation}

\begin{figure}[t]
  \centering
  \includegraphics[width=1.0\columnwidth]{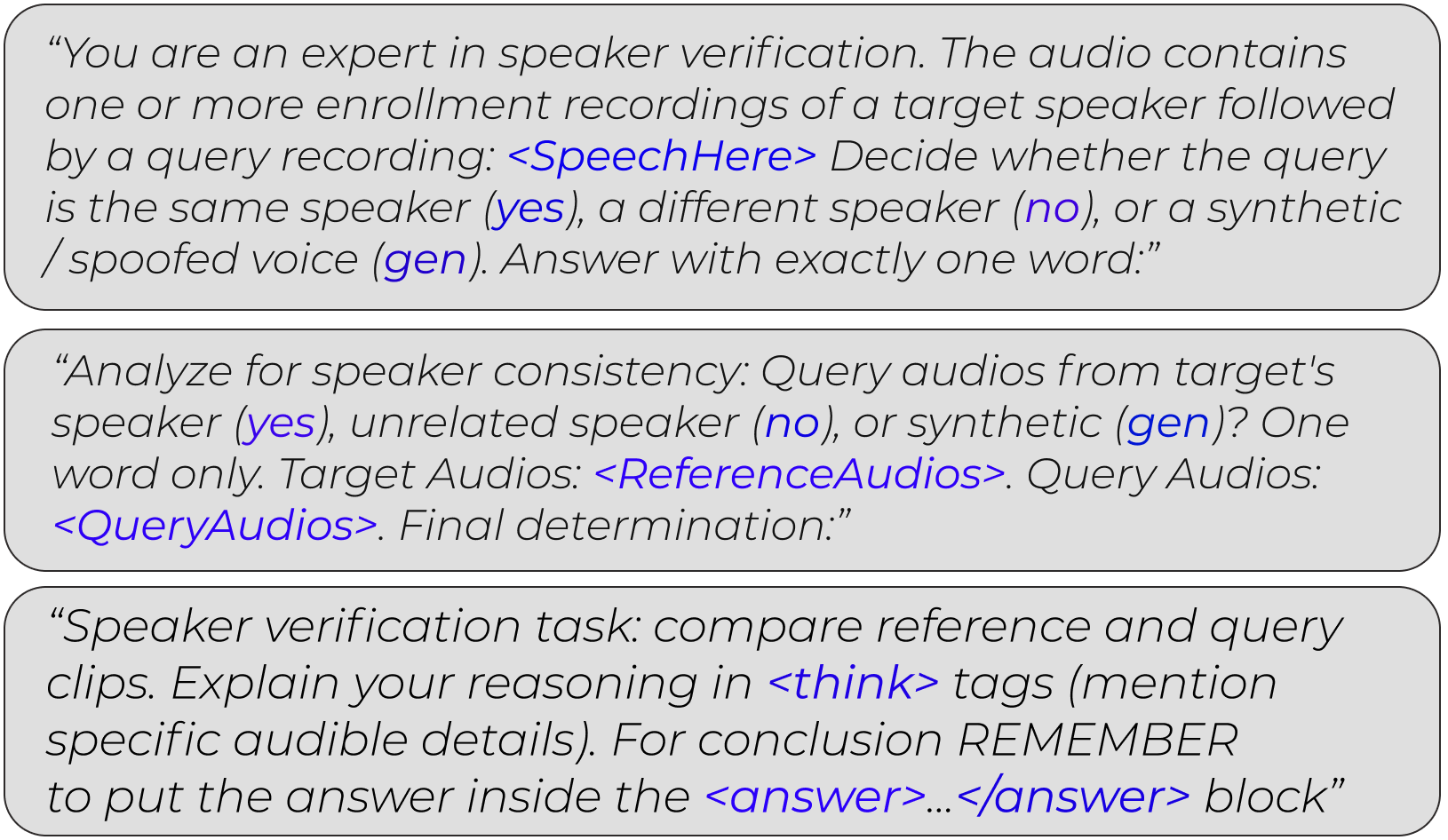}
  \caption{Prompt variants for LALM-based SASV. Top: direct decision over concatenated audio. Middle: multi-audio prompting with separate enrollment/trial tokens. Bottom: reasoning-augmented prompt.}
  \label{fig:prompts}
\end{figure}

Based on results reported in prior work, we expect zero-shot performance on SASV to be weak~\cite{gu2025allm4add,ren2025can}. Thus, for the first branch of our experiments, we consider hard-label only (given input pair, return target/non-target/spoof) supervised-finetuning (SFT) of LALM. The models we use were never adapted to the SASV objective and were mainly trained for speech recognition and audio captioning.
To align the model with the SASV formulation without disrupting its general acoustic knowledge, we apply LoRA parameter-efficient finetuning to the attention layers.
For our experiments, we consider two common LALMs that incorporate two different approaches in multiple audio processing pipelines. Namely, SALMONN~\cite{tang2024salmonn}, while not being trained for multi-audio processing, supports it by sequential concatenation of given audio, particularly, the enrollment and trial utterances at the waveform level with a second of silence and feeds them to the model as a single prompt. Qwen-Audio~\cite{chu2024qwen2audio} instead was trained for multi-audio processing and receives the enrollment and trial utterances as separate audio inputs. The corresponding prompt templates are shown in Figure~\ref{fig:prompts}.

\subsection{Optimization Strategies}

SASV requires the simultaneous optimization of two competing objectives: speaker discrimination and spoof detection. However, an direct SFT can result in a poor trade-off between the predictions of three classes, where, for example, the quality of spoof detection significantly outperforms that of target/non-target discrimination. We therefore attach three supervision heads, each contributing one term to a composite objective: a three-class cross-entropy $\mathcal{L}{1}$ on the SASV decision, an Additive Angular Margin (AAM) loss~\cite{deng2019arcface} $\mathcal{L}{2}$ that improves speaker separability, and a binary cross-entropy $\mathcal{L}{3}$ that sharpens spoof sensitivity.
Both auxiliary losses, $\mathcal{L}{2}$ and $\mathcal{L}{3}$, have their own learnable head, trained jointly with the backbone.
The AAM term $\mathcal{L}{2}$ uses a separate speaker-discriminative head and applies an angular margin between the embedding and the corresponding speaker prototype:

\begin{equation}
\mathcal{L}{2} =
-
\log
\frac{
e^{s\cos(\theta_{i,y_i}+m)}
}{
e^{s\cos(\theta_{i,y_i}+m)}
+
\sum_{j \neq y_i}
e^{s\cos\theta_{i,j}}
}.
\end{equation}

$\mathcal{L}_{2}$ is applied only to bona fide samples: enforcing an identity margin on spoofed utterances would pull synthetic embeddings into the speaker manifold, which leads to degradation of spoof detection. The speaker embeddings fed to the AAM head are obtained from the audio-conditioned
QFormer features, mean-pooled separately
over the enrollment and trial segments; both are supervised.
The spoof term $\mathcal{L}_{3}$ is computed by a linear bona-fide/spoof head, yielding a two-logit classifier supervised by binary cross-entropy. Routing the two auxiliary heads to different representations lets each head specialize in the signal best suited to it. The total objective is a weighted sum of the three terms:
\begin{equation}
\mathcal{L} = \lambda_1\,\mathcal{L}_{1} + \lambda_2\,\mathcal{L}_{2} + \lambda_3\,\mathcal{L}_{3} .
\end{equation}
Beyond the loss formulation, we improve robustness via hard-sample mining. As training proceeds, we progressively concentrate the sampler on the most informative sample pairs: hard-yes pairs (same speaker, low embedding similarity), hard-no pairs (different speakers, high similarity), and hard-spoof pairs (synthetic trials whose
embeddings lie closest to the enrolled speaker). The difficulty of the pair is periodically reinforced as the embedding space evolves, so the model repeatedly confronts boundary cases.

\subsection{Chain-of-Thought SFT}

LALM-based detectors are optimized for the hard label only, leaving their reasoning and interpretability unexploited~\cite{gu2025allm4add}. Nonetheless, reasoning-based supervision has recently been explored for SDD, where chain-of-thought traces~\cite{wang2025speechllm,xu2026holiantispoof,chen2026audio} expose the perceptual cues behind a decision instead of isolated hard-label prediction. A complementary line of research shows that human perceptible cues are predictive of synthesized speech and can be used as supervision for LALMs~\cite{dvirniak2026towards}. 
We carry this perspective over to SASV, where the model must justify not only whether the trial is target, non-target, or synthetic but also provide textual reasoning, explaining and supporting the decision. For this purpose, we enrich a subset of training pairs with explicit reasoning traces and supervise the model to emit a structured output (see Fig.~\ref{fig:prompts}).

We construct the traces automatically so that they scale with the training data. The problem is that due to the poor zero-shot performance of existing LALMs on new tasks, such as SDD and ASV, their cold-start reasoning remains unreliable, and grounding can be mostly hallucinated. To partially mitigate, firstly, we filtered out incorrect annotations, where the annotating LALM produced an incorrect class prediction by the annotating LALM. Moreover, to construct the reasoning traces that contain cues that are grounded in interpretable speaker and paralinguistic attributes (such as age, gender, accent, emotion, and prosody), we decided to explicitly add conventional features, and the following models were used:
CoLMbo~\cite{baali2025colmbo} for speaker profiling, emotion2vec~\cite{ma2024emotion2vec} for emotion, 
and Parselmouth~\cite{jadoul2018parselmouth} for prosodic descriptors.

\begin{figure}[t]
  \centering
  \includegraphics[width=0.46\textwidth]{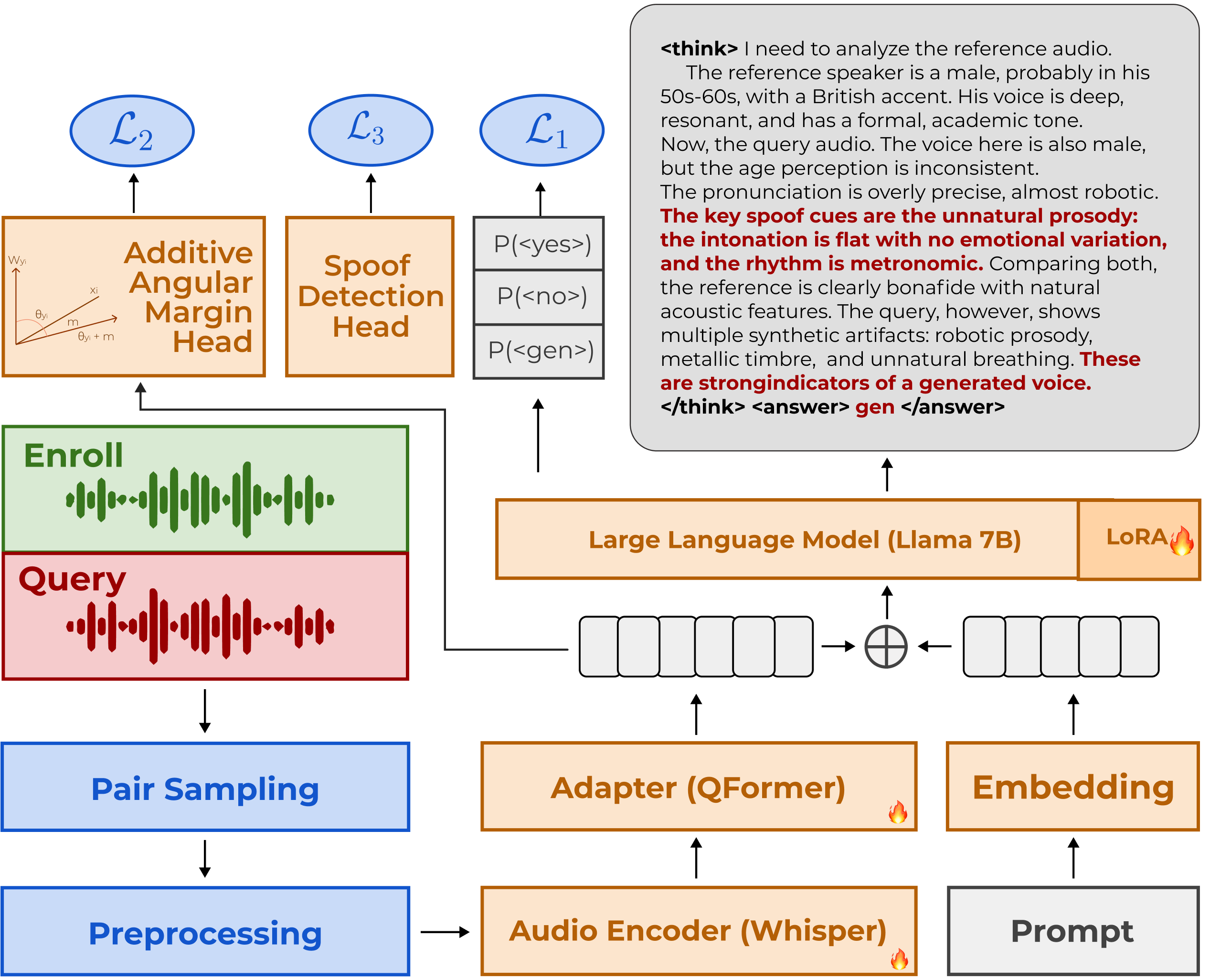}
  \caption{Overview of the proposed SALMONN-based SASV model.}
  \label{fig:arch}
\end{figure}


To further improve CoT quality and diversity by leveraging a diverse set of non-CoT-annotated data,  we apply group relative policy optimization (GRPO)~\cite{shao2024deepseekmath}. The rewards encourage answer correctness and compliance with the required \texttt{<think>}, \texttt{<reasons>}, and \texttt{<answer>} format.

\section{Experimental Setting}

For the base model, we used the SALMONN-7B model, which combines Whisper and BEATS audio encoders via a QFormer adapter with a Vicuna LLM backbone. As an additional model, we use Qwen2-Audio-7B~\cite{chu2024qwen2audio}, which utilized a transformer-based audio encoder. The audio-encoders were primarily set into the training mode, unless stated otherwise. We use LoRA rank $128$, $\alpha=256$, and dropout $0.05$. For optimization, all models are trained with AdamW at a learning rate of $1\times10^{-5}$ with $1000$ warmup steps, an effective batch size of $1024$, for $10$--$20$ epochs. The composite training objective uses weights $\lambda_1{=}1.0$, $\lambda_2{=}0.5$ and $\lambda_3{=}1.0$.

Considering the results from the WildSpoof-2026~\cite{wu2026wildspoof} summary paper, where the worst metrics were obtained on the ASVspoof~5 eval set, we have decided to evaluate the models' performance primarily on this dataset, as one of the most difficult common benchmarks. Training samples were also primarily obtained from the ASVspoof~5 train set. To enrich the genuine side, we add bona fide utterances drawn from VoxCeleb~\cite{Nagrani19voxceleb}, providing additional speaker identities. The VoxCeleb speakers do not overlap with any ASVspoof~5 evaluation speaker, preserving a valid train/test separation. From the combined genuine pool and the ASVspoof~5 audio, we construct enrollment-trial pairs, yielding approximately $1.8$ million pairs spanning $500$ speakers.

All audio is resampled to $16$~kHz, trimmed with voice activity detection, and segment-wise normalized to zero mean and unit variance. Training utterances are randomly cropped or repeat-padded to $6$-$7$~s, and the first $8$~s are used at evaluation. For SALMONN, the enrollment and trial utterances are concatenated into a single waveform with a $1$~-s silence gap inserted between them. We apply random augmentation during training to improve robustness to noise, reverberation, and channel effects. With probability $0.1$, we add background noise at a signal-to-noise ratio (SNR) sampled in $9$--$15$~dB, and with probability $0.07$, we convolve the signal with a recorded room impulse response to simulate reverberation. We additionally use RawBoost~\cite{tak2022rawboost} with probability $0.07$.  Augmentation is drawn independently for each pair's enrollment and trial audio, so the model must verify identity and detect spoofing despite mismatches between the two inputs.

\subsection{Reasoning Annotation}

To obtain reasoning supervision, we first generate candidate traces with a stronger reasoning model, Step-Audio R1~\cite{tian2025step}, used as a cold-start annotator. For each audio pair, we sample three independent traces. However, these traces cannot be used directly. It is well known that LALMs are prone to hallucinations~\cite{tian2025step}. Rather than reasoning toward an answer, they often fit the rationale to a guessed label. As a result, the quality of the generated reasoning is unreliable. This is further confirmed by the poor zero-shot performance. Data filtering is therefore essential for our task. We keep only those traces in which the model is both correct and self-consistent.
A sample is retained only if it satisfies two conditions. First, the predicted decision must be correct in at least two of the three traces, ensuring that the model reaches the right label reliably. Second, the rationales across traces must agree. The cues cited in each trace are required to reach a Jaccard similarity of at least $0.35$, so that the retained samples are supported by a stable set of reasons. Only traces passing both filters form the cold-start reasoning set used for training. After filtering, the reasoning dataset contains approximately $90$K pairs. The generated reasoning traces are relatively long, with an average length of about $500$ tokens, which provides dense supervision.


In practice, the amount of reliable reasoning data that could be generated and retained after filtering was limited. As a result, the reasoning-supervised models did not consistently improve over direct decision-level fine-tuning. This suggests that grounded reasoning for SASV likely requires either substantially larger and higher-quality reasoning annotations or a more explicit audio-evidence verification mechanism.

\begin{table}[tbh]
\centering
\caption{Zero-shot vs.\ adapted LALMs on ASVspoof~5 SASV.}
\label{tab:2}
\begin{tabular}{lll}
\toprule
Model & Setting & Acc.\ (\%) \\
\midrule
Qwen-Audio & zero-shot & 27.71 \\
Qwen-Audio & LoRA SFT   & 85.45 \\
\midrule
SALMONN & zero-shot & 43.75 \\
SALMONN & LoRA SFT     & 85.84 \\
\bottomrule
\end{tabular}
\end{table}

\begin{table*}[tbh]
\centering
\caption{Ablation of adaptation strategies on SALMONN-7B
and comparison with baselines.}
\label{tab:rq3_ablation}
\begin{tabular}{lccccc}
\toprule
Configuration & Acc.\ (\%) & ASV Acc.\ (\%) & CM Acc.\ (\%) & EER\,$\downarrow$ & a-DCF\,$\downarrow$ \\
\midrule
LoRA SFT $\mathcal{L}_{1}$             & 85.84 & 80.52 & 96.47 & 0.13 & 0.26 \\
\;+ $ \mathcal{L}_{2}$ (AAM Head)               & 59.82 & 95.50 & 21.00 & 0.18 & 0.29 \\
\;+ $\mathcal{L}_{3}$  (Spoof Detection Head)   & 86.73 & 90.10 & 80.01 & 0.16 & 0.20 \\
\;+ Hard-sample mining                          & 89.30 & 86.53 & 97.19 & 0.22 & 0.19 \\
\midrule
ECAPA2 + WavLM score fusion~\cite{aliyev24_asvspoof}  
                                     & 79.78 & 81.47 & 77.72 & 0.14 & 0.41\\
ECAPA2 + W2V2-AASIST (decision tree) & 86.67 & 84.07 & 88.46 & 0.13 & 0.32 \\
ECAPA2 + W2V2-AASIST (threshold)     & 86.40 & 87.56  & 85.15 & 0.14 & 0.31 \\
\bottomrule
\end{tabular}
\end{table*}

\subsection{Evaluation Protocol}
We adopt the datasets and the metrics from the WildSpoof and ASVspoof~5 competitions: our model is trained on the train set and evaluated on its evaluation set following the Track~2 protocol under the open condition, which permits external data and pre-trained models. However, it is worth noting the following differences in the evaluation protocol. First, although the evaluation plan specifies three enrollment utterances per claimed identity, all results reported here use only a single enrollment utterance, placing our system under a harder evaluation condition. Second, owing to the size of the full evaluation set and the computational expense of LALMs,
results are computed on a stratified sample of $20$K pairs that preserves its original class distribution.

Turning the model's output into a score for these metrics requires one extra
step, since the model does not emit a continuous score but a decision token.
We therefore recover the score from the posterior over the answer tokens.
Following~\cite{gu2025allm4add}, let $p_{\mathrm{target}}$, $p_{\mathrm{non-target}}$ and $p_{\mathrm{spoof}}$ be the renormalised probabilities (softmax over the corresponding 3 logits) of each decision, we take the SASV score to be the acceptance posterior $s_{\mathrm{sasv}} = p_{\mathrm{target}}$, i.e. the model's confidence that the trial is a target bona fide audio. Sweeping a threshold $\tau$ on $s_{\mathrm{sasv}}$ then yields the curve on which the rate-based metrics are computed.
We report accuracy together with three threshold-based metrics:
the equal error rate (EER) and the primary metric, the minimum
architecture-agnostic detection cost function (min~a-DCF)~\cite{shim2024dcf}.
The a-DCF was designed for integrated SASV systems that output a single score
and is the primary metric of ASVspoof~5 Track~2, whose cost setting we follow:
priors $\pi_{\mathrm{tar}}=0.9405$, $\pi_{\mathrm{non}}=0.0095$,
$\pi_{\mathrm{spf}}=0.05$ and costs $C_{\mathrm{miss}}=1$,
$C_{\mathrm{fa}}=C_{\mathrm{fa,spf}}=10$. The same parameters were utilized in WildSpoof. 
However, for reasoning-oriented models, we report only accuracy, since the final CoT decision is strongly conditioned on the preceding reasoning trace, making the resulting scores poorly calibrated for threshold-based metrics such as EER and min a-DCF.


\subsection{Baseline Competitors}

We compare the adapted LALM systems against three fusion-based baselines that
follow the dominant tandem paradigm in SASV, where an ASV subsystem is combined with a separate countermeasure score. The first competitor is a re-implementation of the power-fusion system from~\cite{aliyev24_asvspoof}, the top-ranked entry in the ASVspoof~5 SASV track, which fuses an ECAPA2- and WavLM-based speaker embedding extractors~\cite{jenthe23ecapa2}. Since the original source code and checkpoints are not publicly available, we re-implement the method and train it under our experimental protocol. We evaluate the standalone model without ensembling or additional post-processing. Consequently, the obtained scores are not directly comparable to the official leaderboard results. The discrepancy is likely attributable to differences in training data scale and implementation details. Nevertheless, under our controlled protocol, the method performs competitively with the remaining baselines and therefore serves as a meaningful reference point. The remaining two baselines fuse an ECAPA2 ASV system with a Wav2Vec2-AASIST CM~\cite{tak2022automatic}. We vary fusion approaches, using a learned decision-tree combiner and a fixed score threshold.

\section{Results}


Table~\ref{tab:2} contrasts the zero-shot behavior of the LALMs. In the zero-shot setting, both backbones perform close to chance, confirming that general-purpose audio understanding does not transfer to SASV out of the box. After LoRA SFT, performance improves, indicating that LALMs can be turned into competitive SASV models, but only after explicit adaptation to the objective.

Table~\ref{tab:rq3_ablation} shows an ablation on the SALMONN-7B based SASV model. LoRA fine-tuning alone already provides a drastic improvement over zero-shot but remains insufficient for speaker discrimination. Adding an AAM loss sharply improves the ASV sub-task, but degrades CM. Adding a dedicated anti-spoofing head trained with a BCE loss recovers the lost CM accuracy without sacrificing the ASV gains, resolving the trade-off. Finally, hard-sample mining provides a boost for accuracy, CM accuracy, and A-DCF.
The competing approaches also achieve strong metrics, confirming that competitive SASV performance is attainable through several distinct routes.
The final system reaches a lower a-DCF than the reproduced fusion baselines under our controlled protocol. Its scores are also in the range of strong published ASVspoof5 systems~\cite{wu2026wildspoof}, although this comparison should be treated only as indicative because the training data and evaluation subset differ from the official challenge setup.

To ensure a fair assessment of reasoning supervision, we additionally train a data-matched label-only SALMONN baseline on the same 90K pairs used for reasoning-oriented training. This control separates the effect of explicit reasoning supervision from the effect of reduced training scale relative to the main 1.8M-pair SALMONN model. We also include a data-matched conventional fusion baseline. The ASV model (ECAPA2) is pretrained, and the CM (W2V2-AASIST) component and the decision-tree fusion module are trained on the same 90K training pairs. The results are presented in Table~\ref{tab:3}. 
All SALMONN variants substantially outperform the conventional fusion approach in terms of overall accuracy. The hard-label model achieves the best overall performance (86.02\%), while CoT-SFT attains the highest ASV accuracy (93.95\%). Although GRPO yields a slightly lower overall accuracy (84.73\%), it provides the best balance between the two subtasks among the reasoning models, achieving the highest CM accuracy (89.04\%) with only a minor reduction in ASV accuracy. It is worth mentioning that the modest performance degradation of the reasoning models may reflect a robustness-performance trade-off, consistent with the analysis in~\cite{nguyen2026analyzing}.


\begin{table}[tbh]
\centering
\caption{Test result comparison of ECAPA2 + W2V2-AASIST (``Conventional fusion'' in the table), hard-label and reasoning-oriented SALMONN training on the same 90K pairs.}
\label{tab:3}
\begin{tabular}{lccc}
\toprule
Method & Acc.\ (\%) & ASV Acc.\ (\%) & CM Acc.\ (\%) \\
\midrule
Conventional fusion    & 72.00 & 83.19 & 61.19 \\
SALMONN hard-label     & 86.02 & 93.83 & 86.78 \\
SALMONN CoT-SFT        & 83.92 & 93.95 & 72.55 \\
SALMONN GRPO           & 84.73 & 90.05 & 89.04 \\
\bottomrule
\end{tabular}
\end{table}

\section{Discussion and Limitations}

The results show that LALMs are not naturally adapted for SASV in the zero-shot scenario. Although these models are pretrained on large-scale audio data and can solve a variety of audio tasks, SASV requires a task-specific decision process. The near-random zero-shot performance suggests that SASV capabilities are not acquired automatically through general audio pretraining. At the same time, LoRA adaptation substantially improves performance. After supervised fine-tuning, the LALM-based systems demonstrate clear gains over the zero-shot setting. 
\begin{tcolorbox}[
    colback=gray!5,
    colframe=gray!45,
    boxrule=0.6pt,
    arc=2mm,
    left=2mm,
    right=2mm,
    top=1mm,
    bottom=1mm
]
\textbf{Takeaway for RQ1:} Common audio-language models can be adapted to perform SASV, but they cannot solve this task directly in a zero-shot setting. To work well, the model needs fine-tuning.
\end{tcolorbox}
The results show that the LALM-based adapted system is competitive with strong conventional SASV solutions. Compared with fusion-based baselines, the final SALMONN-based model achieves the highest overall accuracy and the lowest a-DCF, indicating better end-to-end decision quality for the SASV task. However, conventional systems still achieve better EER values, which indicates stronger threshold-based speaker discrimination in some settings. 

\begin{tcolorbox}[
colback=gray!5,
colframe=gray!45,
boxrule=0.6pt,
arc=2mm,
left=2mm,
right=2mm,
top=1mm,
bottom=1mm
]
\textbf{Takeaway for RQ2:} Adapted LALM-based systems are competitive with conventional SASV baselines and can improve overall SASV decision quality, but classical systems still remain strong in threshold-based speaker discrimination metrics.
\end{tcolorbox}

The ablation study shows that different adaptation components affect speaker verification and spoof detection differently. Fine-tuning already provides a strong performance, especially for spoof detection, but its speaker verification performance limited. Adding an AAM speaker head improves ASV accuracy, demonstrating that speaker-discriminative supervision is useful. At the same time, this modification significantly degrades CM accuracy, revealing a trade-off between speaker discrimination and spoof detection. Adding the spoof-detection head resolves this imbalance. It restores spoof detection performance while maintaining stronger speaker verification than the initial model. Finally, hard-sample mining provides the best overall accuracy and the lowest a-DCF. Although hard mining slightly reduces ASV accuracy and increases EER, it improves application-oriented SASV metrics.
\begin{tcolorbox}[
colback=gray!5,
colframe=gray!45,
boxrule=0.6pt,
arc=2mm,
left=2mm,
right=2mm,
top=1mm,
bottom=1mm
]
\textbf{Takeaway for RQ3:} SASV adaptation involves a trade-off between ASV and CM tasks. Combining multiple adaptation strategies, such as dedicated speaker and spoof heads and hard-sample mining, helps balance this trade-off and leads to stronger results.
\end{tcolorbox}

LALMs under-perform in terms of EER compared to conventional methods. This discrepancy can be attributed to the fact that token-level probabilities are not trained as calibrated verification scores. The softmax over a small set of decision tokens provides a convenient proxy score, but it is affected by instruction-following behavior, priors, prompt formatting, and the generative decoding context. Therefore, improvements in hard-label decision accuracy may not translate into better threshold-based metrics such as EER.

\begin{tcolorbox}[
colback=gray!5,
colframe=gray!45,
boxrule=0.6pt,
arc=2mm,
left=2mm,
right=2mm,
top=1mm,
bottom=1mm
]
\textbf{Takeaway (RQ4):} Explicit reasoning methods do not improve overall performance over hard-label adaptation, and they still outperform the conventional fusion baseline. Their main benefit is the ability to produce interpretable reasoning traces, which might be useful for analyzing model decisions.
\end{tcolorbox}

However, proposed method has several limitations. Despite their strong performance, LALM-based SASV systems require substantially higher computational resources than conventional fusion models. In addition, although the generated reasoning traces appear plausible, they have not yet been systematically validated. Therefore, we leave human evaluation of reasoning quality for future work.

\section{Conclusion}

In this work, we studied whether Large Audio Language Models can be adapted for spoofing-aware speaker verification. We formulated SASV as a pair-based audio task and evaluated LALMs under zero-shot scenario, LoRA adaptation, and different training strategies. Our ablation study shows that SASV adaptation requires balancing two competing objectives: speaker discrimination and spoof detection. LoRA fine-tuning provides a strong starting point, but additional speaker- and spoof-oriented supervision is needed to improve the final quality. In particular, the combination of a different training strategies  leads to the strongest overall performance in terms of accuracy and min a-DCF. An additional improvement arises from hard-sample mining.
Overall, reasoning-oriented training does not uniformly improve hard-label supervision. GRPO improves spoof-awareness, while CoT-SFT preserves strong ASV performance.


\section*{Acknowledgment}
AI models were used for the creation of reasoning traces, for grammar correction and for text polishing and shortening.

\bibliographystyle{IEEEtran}
\bibliography{mybib}

@article{nguyen2026analyzing,
  title={Analyzing Reasoning Shifts in Audio Deepfake Detection under Adversarial Attacks: The Reasoning Tax versus Shield Bifurcation},
  author={Nguyen, Binh and Le, Thai},
  journal={arXiv preprint arXiv:2601.03615},
  year={2026}
}

@misc{shao2024deepseekmath,
  author = {Zhihong Shao and others},
  title = {DeepSeekMath: Pushing the Limits of Mathematical Reasoning in Open Language Models},
  journal = {CoRR},
  volume = {abs/2402.03300},
  year = {2024},
  url = {https://arxiv.org/abs/2402.03300},
}

@inproceedings{zeng2024spoofing,
  title={Spoofing-aware speaker verification robust against domain and channel mismatches},
  author={Zeng, Chang and Miao, Xiaoxiao and Wang, Xin and Cooper, Erica and Yamagishi, Junichi},
  booktitle={2024 IEEE Spoken Language Technology Workshop (SLT)},
  pages={1150--1157},
  year={2024},
  organization={IEEE}
}

@article{dinkel2024scaling,
  title={Scaling up masked audio encoder learning for general audio classification},
  author={Dinkel, Heinrich and Yan, Zhiyong and Wang, Yongqing and Zhang, Junbo and Wang, Yujun and Wang, Bin},
  journal={arXiv preprint arXiv:2406.06992},
  year={2024}
}

@article{peng2025but,
  title={BUT Systems for WildSpoof Challenge: SASV in the Wild},
  author={Peng, Junyi and Li, Jin and Rohdin, Johan and Zhang, Lin and Hlav{\'a}{\v{c}}ek, Miroslav and Plchot, Oldrich},
  journal={arXiv preprint arXiv:2512.12851},
  year={2025}
}

@article{jung2025spoofceleb,
  title={Spoofceleb: Speech deepfake detection and sasv in the wild},
  author={Jung, Jee-weon and Wu, Yihan and Wang, Xin and Kim, Ji-Hoon and Maiti, Soumi and Matsunaga, Yuta and Shim, Hye-jin and Tian, Jinchuan and Evans, Nicholas and Chung, Joon Son and others},
  journal={IEEE Open Journal of Signal Processing},
  year={2025},
  publisher={IEEE}
}

@article{zhang2026overview,
  title={Overview of ESDD2: Environment-Aware Speech and Sound Deepfake Detection Challenge},
  author={Zhang, Xueping and Yin, Han and Xiao, Yang and Zhang, Lin and Dang, Ting and Das, Rohan Kumar and Li, Ming},
  journal={arXiv preprint arXiv:2606.10791},
  year={2026}
}

@article{luong2026radar,
  title={RADAR Challenge 2026: Robust Audio Deepfake Recognition under Media Transformations},
  author={Luong, Hieu-Thi and Liu, Xuechen and Kukanov, Ivan and Chai, Zheng Xin and Lee, Kong Aik},
  journal={arXiv preprint arXiv:2605.09568},
  year={2026}
}

@article{xie2026add,
  title={AT-ADD: All-Type Audio Deepfake Detection Challenge Evaluation Plan},
  author={Xie, Yuankun and Cheng, Haonan and Zhou, Jiayi and Guo, Xiaoxuan and Wang, Tao and Liu, Jian and Wang, Weiqiang and Fu, Ruibo and Wang, Xiaopeng and Huang, Hengyan and others},
  journal={arXiv preprint arXiv:2604.08184},
  year={2026}
}

@article{zhu2026alethia,
  title={Alethia: A Foundational Encoder for Voice Deepfakes},
  author={Zhu, Yi and Dwivedi, Brahmi and Raghuram, Jayaram and Koppisetti, Surya},
  journal={arXiv preprint arXiv:2605.00251},
  year={2026}
}

@article{dvirniak2026towards,
  title={Towards Robust Speech Deepfake Detection via Human-Inspired Reasoning},
  author={Dvirniak, Artem and Kushnir, Evgeny and Tarasov, Dmitrii and Iudin, Artem and Kiriukhin, Oleg and Pautov, Mikhail and Korzh, Dmitrii and Rogov, Oleg Y},
  journal={arXiv preprint arXiv:2603.10725},
  year={2026}
}

@article{tan2026integrated,
  title={Integrated Spoofing-Robust Automatic Speaker Verification via a Three-Class Formulation and LLR},
  author={Tan, Kai and Zhang, Lin and Zhang, Ruiteng and Rohdin, Johan and Garc{\'\i}a-Perera, Leibny Paola and Cai, Zexin and Khudanpur, Sanjeev and Wiesner, Matthew and Andrews, Nicholas},
  journal={arXiv preprint arXiv:2603.13780},
  year={2026}
}

@article{kurnaz2026joint,
  title={Joint optimization of speaker and spoof detectors for spoofing-robust automatic speaker verification},
  author={Kurnaz, O{\u{g}}uzhan and Mishra, Jagabandhu and Kinnunen, Tomi H and Hanil{\c{c}}i, Cemal},
  journal={IEEE Transactions on Audio, Speech and Language Processing},
  year={2026},
  publisher={IEEE}
}

@inproceedings{wang2022dku,
  title={The DKU-OPPO System for the 2022 Spoofing-Aware Speaker Verification Challenge},
  author={Wang, Xingming and Qin, Xiaoyi and Wang, Yikang and Xu, Yunfei and Li, Ming},
  booktitle={Proc. Interspeech 2022},
  pages={4396--4400},
  year={2022}
}

@inproceedings{heo2022two,
  title={Two Methods for Spoofing-Aware Speaker Verification: Multi-Layer Perceptron Score Fusion Model and Integrated Embedding Projector},
  author={Heo, Jungwoo and Kim, Ju-Ho and Shin, Hyun-seo},
  booktitle={Proc. Interspeech 2022},
  pages={2878--2882},
  year={2022}
}

@inproceedings{shim2022baseline,
  title={Baseline Systems for the First Spoofing-Aware Speaker Verification Challenge: Score and Embedding Fusion},
  author={Shim, Hye-jin and Tak, Hemlata and Liu, Xuechen and Heo, Hee-Soo and Jung, Jee-weon and Chung, Joon Son and Chung, Soo-Whan and Yu, Ha-Jin and Lee, Bong-Jin and Todisco, Massimiliano and others},
  booktitle={Proc. Odyssey 2022},
  pages={330--337},
  year={2022}
}

@article{desplanques2020ecapa,
  title={ECAPA-TDNN: Emphasized Channel Attention, Propagation and Aggregation in TDNN Based Speaker Verification},
  author={Desplanques, Brecht and Thienpondt, Jenthe and Demuynck, Kris},
  journal={Interspeech 2020},
  year={2020},
  publisher={ISCA}
}

@article{nam2026speakerllm,
  title={SpeakerLLM: A Speaker-Specialized Audio-LLM for Speaker Understanding and Verification Reasoning},
  author={Nam, KiHyun and Heo, Jungwoo and Bae, Siu and Yu, Ha-Jin and Chung, Joon Son},
  journal={arXiv preprint arXiv:2605.15044},
  year={2026}
}

@article{ren2025can,
  title={Can Audio Large Language Models Verify Speaker Identity?},
  author={Ren, Yiming and Xu, Xuenan and Li, Baoxiang and Wang, Shuai and Zhang, Chao},
  journal={arXiv preprint arXiv:2509.19755},
  year={2025}
}

@inproceedings{Alenin2022A,
  author={Alenin, Alexander and Torgashov, Nikita and Okhotnikov, Anton and Makarov, Rostislav and Yakovlev, Ivan},
  title={A Subnetwork Approach for Spoofing Aware Speaker Verification},
  year=2022,
  booktitle={Proc. Interspeech (submitted)}
}

@article{jung2022sasv,
  title={SASV 2022: The First Spoofing-Aware Speaker Verification Challenge},
  author={Jung, Jee-weon and Tak, Hemlata and Shim, Hye-jin and Heo, Hee-Soo and Lee, Bong-Jin and Chung, Soo-Whan and Yu, Ha-Jin and Evans, Nicholas and Kinnunen, Tomi},
  journal={Interspeech 2022},
  pages={2893--2897},
  year={2022},
  publisher={ISCA}
}

@inproceedings{wu2026wildspoof,
  title={WildSpoof: advancing in-the-wild data in Text-to-Speech generation and Spoofing-aware automatic speaker verification},
  author={Wu, Yihan and Jung, Jee-Weon and Shim, Hye-Jin and Cheng, Xin and Wang, Xin},
  booktitle={ICASSP 2026-2026 IEEE International Conference on Acoustics, Speech and Signal Processing (ICASSP)},
  pages={21922--21924},
  year={2026},
  organization={IEEE}
}

@article{du2025cosyvoice,
  title={Cosyvoice 3: Towards in-the-wild speech generation via scaling-up and post-training},
  author={Du, Zhihao and Gao, Changfeng and Wang, Yuxuan and Yu, Fan and Zhao, Tianyu and Wang, Hao and Lv, Xiang and Wang, Hui and Ni, Chongjia and Shi, Xian and others},
  journal={arXiv preprint arXiv:2505.17589},
  year={2025}
}

@article{chen2026audio,
  title={Audio Language Model for Deepfake Detection Grounded in Acoustic Chain-of-Thought},
  author={Chen, Runkun and Fang, Yixiong and Chang, Pengyu and Li, Yuante and Baali, Massa and Raj, Bhiksha},
  journal={arXiv preprint arXiv:2603.28021},
  year={2026}
}

@article{baali2025colmbo,
  title={CoLMbo: Speaker Language Model for Descriptive Profiling},
  author={Baali, Massa and Han, Shuo and Hannan, Syed Abdul and Samal, Purusottam and Singh, Karanveer and Deshmukh, Soham and Singh, Rita and Raj, Bhiksha},
  journal={arXiv preprint arXiv:2506.09375},
  year={2025}
}

@article{chen2022wavlm,
  title={Wavlm: Large-scale self-supervised pre-training for full stack speech processing},
  author={Chen, Sanyuan and Wang, Chengyi and Chen, Zhengyang and Wu, Yu and Liu, Shujie and Chen, Zhuo and Li, Jinyu and Kanda, Naoyuki and Yoshioka, Takuya and Xiao, Xiong and others},
  journal={IEEE Journal of Selected Topics in Signal Processing},
  volume={16},
  number={6},
  pages={1505--1518},
  year={2022},
  publisher={IEEE}
}

@inproceedings{jiang2024qformer,
  title={QFormer: An Efficient Quaternion Transformer for Image Denoising.},
  author={Jiang, Bo and Lu, Yao and Lu, Guangming and Zhang, Bob},
  booktitle={IJCAI},
  pages={4237--4245},
  year={2024}
}

@article{grattafiori2024llama,
  title={The llama 3 herd of models},
  author={Grattafiori, Aaron and Dubey, Abhimanyu and Jauhri, Abhinav and Pandey, Abhinav and Kadian, Abhishek and Al-Dahle, Ahmad and Letman, Aiesha and Mathur, Akhil and Schelten, Alan and Vaughan, Alex and others},
  journal={arXiv preprint arXiv:2407.21783},
  year={2024}
}

@inproceedings{
  tang2024salmonn,
  title={{SALMONN}: Towards Generic Hearing Abilities for Large Language Models},
  author={Changli Tang and Wenyi Yu and Guangzhi Sun and Xianzhao Chen and Tian Tan and Wei Li and Lu Lu and Zejun MA and Chao Zhang},
  booktitle={The Twelfth International Conference on Learning Representations},
  year={2024},
  url={https://openreview.net/forum?id=14rn7HpKVk}
}

@inproceedings{gu2025allm4add,
  title={Allm4add: Unlocking the capabilities of audio large language models for audio deepfake detection},
  author={Gu, Hao and Yi, Jiangyan and Wang, Chenglong and Tao, Jianhua and Lian, Zheng and He, Jiayi and Ren, Yong and Chen, Yujie and Wen, Zhengqi},
  booktitle={Proceedings of the 33rd ACM International Conference on Multimedia},
  pages={11736--11745},
  year={2025}
}

@article{wang2024asvspoof,
  title={ASVspoof 5: Crowdsourced speech data, deepfakes, and adversarial attacks at scale},
  author={Wang, Xin and Delgado, H{\'e}ctor and Tak, Hemlata and Jung, Jee-weon and Shim, Hye-jin and Todisco, Massimiliano and Kukanov, Ivan and Liu, Xuechen and Sahidullah, Md and Kinnunen, Tomi and others},
  journal={arXiv preprint arXiv:2408.08739},
  year={2024}
}

@article{yi2023add,
	title        = {Add 2023: the second audio deepfake detection challenge},
	author       = {Yi, Jiangyan and others},
	year         = 2023,
	journal      = {arXiv preprint arXiv:2305.13774}
}

@inproceedings{tak2021end,
  title={End-to-end anti-spoofing with rawnet2},
  author={Tak, Hemlata and Patino, Jose and Todisco, Massimiliano and Nautsch, Andreas and Evans, Nicholas and Larcher, Anthony},
  booktitle={ICASSP 2021-2021 IEEE International Conference on Acoustics, Speech and Signal Processing (ICASSP)},
  pages={6369--6373},
  year={2021},
  organization={IEEE}
}

@inproceedings{jung2022aasist,
  title={Aasist: Audio anti-spoofing using integrated spectro-temporal graph attention networks},
  author={Jung, Jee-weon and Heo, Hee-Soo and Tak, Hemlata and Shim, Hye-jin and Chung, Joon Son and Lee, Bong-Jin and Yu, Ha-Jin and Evans, Nicholas},
  booktitle={ICASSP 2022-2022 IEEE international conference on acoustics, speech and signal processing (ICASSP)},
  pages={6367--6371},
  year={2022},
  organization={IEEE}
}

@inproceedings{ding2023samo,
  title={SAMO: Speaker attractor multi-center one-class learning for voice anti-spoofing},
  author={Ding, Siwen and Zhang, You and Duan, Zhiyao},
  booktitle={ICASSP 2023-2023 IEEE International Conference on Acoustics, Speech and Signal Processing (ICASSP)},
  pages={1--5},
  year={2023},
  organization={IEEE}
}

@article{tak2022automatic,
  title={Automatic speaker verification spoofing and deepfake detection using wav2vec 2.0 and data augmentation},
  author={Tak, Hemlata and Todisco, Massimiliano and Wang, Xin and Jung, Jee-weon and Yamagishi, Junichi and Evans, Nicholas},
  journal={arXiv preprint arXiv:2202.12233},
  year={2022}
}

@article{tang2023salmonn,
  title={Salmonn: Towards generic hearing abilities for large language models},
  author={Tang, Changli and Yu, Wenyi and Sun, Guangzhi and Chen, Xianzhao and Tan, Tian and Li, Wei and Lu, Lu and Ma, Zejun and Zhang, Chao},
  journal={arXiv preprint arXiv:2310.13289},
  year={2023}
}

@article{guo2025deepseek,
  title={Deepseek-r1: Incentivizing reasoning capability in llms via reinforcement learning},
  author={Guo, Daya and Yang, Dejian and Zhang, Haowei and Song, Junxiao and Wang, Peiyi and Zhu, Qihao and Xu, Runxin and Zhang, Ruoyu and Ma, Shirong and Bi, Xiao and others},
  journal={arXiv preprint arXiv:2501.12948},
  year={2025}
}

@article{yang2025sakura,
  title={Sakura: On the multi-hop reasoning of large audio-language models based on speech and audio information},
  author={Yang, Chih-Kai and Ho, Neo and Piao, Yen-Ting and Lee, Hung-yi},
  journal={arXiv preprint arXiv:2505.13237},
  year={2025}
}

@article{tian2025step,
  title={Step-Audio-R1 Technical Report},
  author={Tian, Fei and Zhang, Xiangyu Tony and Zhang, Yuxin and Zhang, Haoyang and Li, Yuxin and Liu, Daijiao and Deng, Yayue and Wu, Donghang and Chen, Jun and Zhao, Liang and others},
  journal={arXiv preprint arXiv:2511.15848},
  year={2025}
}

@article{wang2025speechllm,
  title={SpeechLLM-as-Judges: Towards General and Interpretable Speech Quality Evaluation},
  author={Wang, Hui and Zhao, Jinghua and Yang, Yifan and Liu, Shujie and Chen, Junyang and Zhang, Yanzhe and Zhao, Shiwan and Li, Jinyu and Zhou, Jiaming and Sun, Haoqin and others},
  journal={arXiv preprint arXiv:2510.14664},
  year={2025}
}

@article{xu2026holiantispoof,
  title={HoliAntiSpoof: Audio LLM for Holistic Speech Anti-Spoofing},
  author={Xu, Xuenan and Ren, Yiming and Liu, Liwei and Wu, Wen and Li, Baoxiang and Lu, Chaochao and Wang, Shuai and Zhang, Chao},
  journal={arXiv preprint arXiv:2602.04535},
  year={2026}
}

@article{casanova2024xtts,
  title={Xtts: a massively multilingual zero-shot text-to-speech model},
  author={Casanova, Edresson and Davis, Kelly and G{\"o}lge, Eren and G{\"o}knar, G{\"o}rkem and Gulea, Iulian and Hart, Logan and Aljafari, Aya and Meyer, Joshua and Morais, Reuben and Olayemi, Samuel and others},
  journal={arXiv preprint arXiv:2406.04904},
  year={2024}
}

@article{hu2022lora,
  title={Lora: Low-rank adaptation of large language models.},
  author={Hu, Edward J and Shen, Yelong and Wallis, Phillip and Allen-Zhu, Zeyuan and Li, Yuanzhi and Wang, Shean and Wang, Liang and Chen, Weizhu and others},
  journal={Iclr},
  volume={1},
  number={2},
  pages={3},
  year={2022}
}

@article{chu2024qwen2audio,
  title   = {Qwen2-Audio Technical Report},
  author  = {Chu, Yunfei and Xu, Jin and Yang, Qian and others},
  journal = {arXiv preprint arXiv:2407.10759},
  year    = {2024}
}

@inproceedings{deng2019arcface,
  title     = {{ArcFace}: Additive Angular Margin Loss for Deep Face Recognition},
  author    = {Deng, Jiankang and Guo, Jia and Xue, Niannan and Zafeiriou, Stefanos},
  booktitle = {IEEE/CVF Conference on Computer Vision and Pattern Recognition (CVPR)},
  pages     = {4690--4699},
  year      = {2019}
}

@inproceedings{ma2024emotion2vec,
  title     = {emotion2vec: Self-Supervised Pre-Training for Speech Emotion Representation},
  author    = {Ma, Ziyang and Zheng, Zhisheng and Ye, Jiaxin and Li, Jinchao and
               Gao, Zhifu and Zhang, Shiliang and Chen, Xie},
  booktitle = {Findings of the Association for Computational Linguistics: ACL 2024},
  pages     = {15747--15760},
  year      = {2024}
}

@article{jadoul2018parselmouth,
  title   = {Introducing {Parselmouth}: A {Python} interface to {Praat}},
  author  = {Jadoul, Yannick and Thompson, Bill and de Boer, Bart},
  journal = {Journal of Phonetics},
  volume  = {71},
  pages   = {1--15},
  year    = {2018}
}

@Article{Nagrani19voxceleb,
    author       = "Arsha Nagrani and Joon~Son Chung and Weidi Xie and Andrew Zisserman",
    title        = "Voxceleb: Large-scale speaker verification in the wild",
    journal      = "Computer Science and Language",
    year         = "2019",
    publisher    = "Elsevier",
}

@article{shim2024dcf,
  title={a-DCF: an architecture agnostic metric with application to spoofing-robust speaker verification},
  author={Shim, Hye-jin and Jung, Jee-weon and Kinnunen, Tomi and Evans, Nicholas and Bonastre, Jean-Fran{\c{c}}ois and Lapidot, Itshak},
  journal={arXiv preprint arXiv:2403.01355},
  year={2024}
}

@inproceedings{tak2022rawboost,
  title={Rawboost: A raw data boosting and augmentation method applied to automatic speaker verification anti-spoofing},
  author={Tak, Hemlata and Kamble, Madhu and Patino, Jose and Todisco, Massimiliano and Evans, Nicholas},
  booktitle={ICASSP 2022-2022 IEEE International Conference on Acoustics, Speech and Signal Processing (ICASSP)},
  pages={6382--6386},
  year={2022},
  organization={IEEE}
}

@inproceedings{jenthe23ecapa2,
  title={ECAPA2: A hybrid neural network architecture and training strategy for robust speaker embeddings},
  author={Thienpondt, Jenthe and Demuynck, Kris},
  booktitle={2023 IEEE automatic speech recognition and understanding workshop (ASRU)},
  pages={1--8},
  year={2023},
  organization={IEEE}
}

@inproceedings{aliyev24_asvspoof,
  title     = {{Intema system description for the ASVspoof5 Challenge: power weighted score fusion}},
  author    = {Ali Aliyev and Alexander Kondratev},
  year      = {2024},
  booktitle = {{The Automatic Speaker Verification Spoofing Countermeasures Workshop (ASVspoof 2024)}},
  pages     = {152--157},
  doi       = {10.21437/ASVspoof.2024-22},
}

\end{document}